\documentclass[usenatbib]{mnras}
\usepackage{newtxtext,newtxmath}
\usepackage[T1]{fontenc}
\usepackage{ae,aecompl}
\usepackage{color}
\usepackage{graphicx,latexsym,amssymb,geometry,pdflscape,pdfpages,amsmath}
\usepackage{natbib}
\usepackage[modulo]{lineno}
\usepackage{authblk}
\usepackage{dblfloatfix}
\usepackage{multirow}
\usepackage{arydshln} 

\usepackage{natbib}

\newcommand\arcm{\ensuremath{^\prime}}

\newcommand{\fermi}{{\textit {Fermi}}}

\newcommand{\snr}{SNR\,G24.7+0.6}
\newcommand{\hess}{HESS\,J1837--069}
\newcommand{\mgc}{MAGIC\,J1835--069}
\newcommand{\mgcS}{MAGIC\,J1837--073}

\title[Observations of FoV of SNR G24.7+0.6 by MAGIC]{Discovery of TeV $\gamma$-ray emission from the neighbourhood of the supernova remnant G24.7+0.6 by MAGIC}

\author[MAGIC Collaboration]
{\parbox{\textwidth}{MAGIC Collaboration,
V.~A.~Acciari$^{1}$,
S.~Ansoldi$^{2,20}$,
L.~A.~Antonelli$^{3}$,
A.~Arbet Engels$^{4}$,
C.~Arcaro$^{5}$,
D.~Baack$^{6}$,
A.~Babi\'c$^{7}$,
B.~Banerjee$^{8}$,
P.~Bangale$^{9}$,
U.~Barres de Almeida$^{9,10}$
% ...

and
E.~de O\~na Wilhelmi$^{27,28,29}$\thanks{},
D.~F.~Torres$^{27,28,30}$,
R.~Zanin$^{31}$\thanks{},\\

(Affiliations can be found after the references)
}}

\begin{document}

\maketitle
\clearpage


\label{firstpage}
\pagerange{\pageref{firstpage}--\pageref{lastpage}}



\begin{abstract}
SNR G24.7+0.6 is a 9.5 kyrs radio and $\gamma$-ray supernova remnant evolving in a dense medium. In the GeV regime, SNR G24.7+0.6 (3FHL\,J1834.1--0706e/FGES\,J1834.1--0706) shows a hard spectral index ($\Gamma$$\sim$2) up to $200$\,GeV, which makes it a good candidate to be observed with Cherenkov telescopes such as MAGIC.
We observed the field of view of \snr\ with the MAGIC telescopes for a total of 31 hours. We detect very high energy $\gamma$-ray emission from an extended source located 0.34\degr\ away from the center of the radio SNR. The new source, named \mgc\ is detected up to 5\,TeV, and its spectrum is well-represented by a power-law function with spectral index of $2.74 \pm 0.08$. The complexity of the region makes the identification of the origin of the very-high energy emission difficult, however the spectral agreement with the LAT source and overlapping position at less than 1.5$\sigma$ point to a common origin. We analysed 8 years of \fermi-LAT data to extend the spectrum of the source down to 60\,MeV. \fermi-LAT and MAGIC spectra overlap within errors and the global broad band spectrum is described by a power-law with exponential cutoff at $1.9\pm0.5$\,TeV. The detected $\gamma$-ray emission can be interpreted as the results of proton-proton interaction between the supernova and the CO-rich surrounding. 
\end{abstract}



\section{Introduction}
Composite supernova remnants (SNRs) are known to accelerate particles to very high energies (VHE), up to hundreds of TeV or beyond \citep{2007ApJ...664L..87A, 2012A&A...541A..13A}, in their expanding shocks and/or the relativistic wind surrounding the left-over, energetic pulsar. Both leptonic and hadronic non-thermal mechanisms produce $\gamma$-ray emission that extends from a few hundreds of MeV to tens of TeV.
This radiation can be generated by the interaction of relativistic electrons scattering off low-energy photon fields, and/or by pion production and decay from direct inelastic collisions of ultrarelativistic protons with target protons of the interstellar medium \citep{longair}. 

\snr\ is a $0.5^{\circ} \times 0.25^{\circ}$ center-filled SNR located at a distance of $\sim$5 kpc 
\citep{1984A&A...133L...4R,1989A&A...216..193L}. It was discovered at radio frequencies as a couple of incomplete shells 
centered at $\rm{RA}_{J2000}=278.57^{\circ}$ and $\rm{DEC}_{J2000}=-7.09^{\circ}$, and a linearly polarized central core with a flat radio spectrum of $\alpha=-0.17$ \citep{1984A&A...133L...4R}, 
indicating the presence of a central pulsar wind nebula (PWN) powered by an undetected pulsar. With an estimated age of 9.5 kyrs 
\citep{1989A&A...216..193L} it belongs to the class of middle-aged SNRs interacting with molecular clouds (MC) as suggested by 
observations in the infrared (IR) energy band and by the detection of $^{13}$CO J=1-0 line at 110 GHz (Galactic Ring Survey, 
\citealt{2006ApJS..163..145J}). 
\citet{2008BAAA...51..209P,2012A&A...538A..14P} discovered several molecular structures, including a molecular arm extending into the center of the SNR and two clouds bordering the remnant. An observation using 
VLA also revealed several ultracompact H {\small II} regions within the SNR. The presence of many young stellar objects in the 
interaction region between the SNR and the MC \citep{2010BAAA...53..221P} also suggests that the large activity related to the SN in the region might be triggering stellar formation.

In X-rays, the SNR was observed with the Einstein Observatory. Although not included in the Einstein catalog of SNRs 
\citep{1990ApJS...73..781S}, \cite{1989A&A...216..193L} derived a flux over the entire SNR region of $\left(3.9 \pm 0.9\right)
\times10^{-13}$\,erg\,cm$^{-2}$\,s$^{-1}$.
The same data yield an upper limit (UL) to a 
differential flux under the assumption of a point source (<2\arcm\ diameter) and extended (circle of 8\arcm\ radius) emission of 
$<1\times10^{-12}$\,erg\,cm$^{-2}$\,s$^{-1}$ and $<3\times10^{-12}$\,erg\,cm$^{-2}$\,s$^{-1}$, respectively. No pulsar or 
PWN has been found yet, although an attempt was done with \textit{XMM}-Newton (OBS. ID:0301880301, PI: O. 
Kargaltsev). Unfortunately, a strong flare in the field of view (FoV) affected the observation, reducing the useful exposure to only 3.5\,ks and limiting the 
sensitivity of the observations.

At GeV energies, \fermi-LAT \citep{2009ApJ...697.1071A} proved to be efficient in detecting SNRs  \citep{2015ApJS..218...23A,2016ApJS..224....8A,2016ApJS..222....5A}. Above 100 MeV, two populations of SNRs seem to be emerging: a 
population of young, X-ray bright, SNRs \citep{2011ApJ...734...28A,2011ApJ...740L..51T} and a second one including evolved 
GeV-bright SNRs, interacting with MCs \citep{2012A&A...546A..21R,2010ApJ...712..459A}. \snr\ belongs to the 
latter group. Although initially associated with the pointlike source 3FGL\,J1833.9--0711, \snr\ appears in the first \fermi\ SNR 
catalog \citep{2016ApJS..224....8A} as an extended source (TS$_{\rm ext}$ = 24.89) with a gaussian morphology of radius 
$0.25^{\circ}\pm0.04_{\rm stat}^{\circ}\pm0.12_{\rm sys}^{\circ}$ centered at $\rm{RA}_{J2000}=278.60^{\circ}\pm0.03_{\rm 
stat}^{\circ}\pm0.1_{\rm sys}^{\circ}$ and $\rm{DEC}_{J2000}=-7.17^{\circ}\pm0.03_{\rm stat}^{\circ}\pm0.03_{\rm sys}^{\circ}$. 
The \fermi-LAT extension is compatible with the radio size, 
but offset by $0.08^{\circ}$ towards the star-forming region G24.73+0.69. Its extension at energies larger than 10 GeV was 
confirmed by the presence of the SNR in both the catalog of extended sources in the Galactic plane (FGES, \citet{2017ApJ...843..139A}) and the third catalog of hard $Fermi$-LAT sources (3FHL, \citet{2017ApJS..232...18A}). \snr\ has been, in fact, identified 
with FGES\,J1834.1--0706 and 3FHL\,J1834.1--0706e. The 3FHL tag confirms the hard spectral nature of the source, thus a 
potential VHE $\gamma$-ray emitter. The spectral results of the sources identified with the \snr\ are all compatible within each 
other showing that the energy spectrum is well-represented with a power-law function of index $2.2$. We take as reference from 
now on the spectral results in the FGES catalog \citep{2017ApJ...843..139A}: a photon index of $2.28\pm0.14$ and an 
integral flux from 10 GeV to 2 TeV of $(5.37\pm0.66)\times10^{-10}$\, erg cm$^{-2}$ s$^{-1}$. 

Above $\sim$500\,GeV, the region was covered by the H.E.S.S. Galactic Plane Survey (HGPS, \citealt{2015ICRC...34..773D}). The 
HGPS shows a large and bright source, dubbed \hess\ \citep{2005Sci...307.1938A,2006ApJ...636..777A}, located $
0.9^{\circ}$ away (at $\rm{RA}_{J2000}=279.41^{\circ}$ and $\rm{DEC}_{J2000}=-6.95^{\circ}$) from \snr. \hess\ has 
an elliptical extension of $0.12^{\circ}\pm 0.02^{\circ}$ and $0.05^{\circ}\pm 0.02^{\circ}$ (with an orientation angle of the semi-major axis of $
\omega=149^{\circ}\pm10^{\circ}$ counterclockwise with respect to the positive Galactic latitude axis) 
at energies above 200\,GeV. The power-law spectrum of \hess\ exhibits a photon index of $2.27\pm0.06$ and an integral flux 
above 200 GeV of $(30.4\pm1.6)\times10^{-12}$cm$^{-2}$s$^{-1}$. \hess\ has been classified as a PWNe, associated to the pulsar PSR\,J1838--065 (or AX\,J1838.0--0655) \citep{2008ApJ...681..515G}. Deeper observations of the region around \hess\ \citep{2008AIPC.1085..320M} led to a more detailed morphological analysis resulting in a new position of \hess\, offset $0.05^{\circ}$ from the initial report at $\rm{RA}_{J2000}=279.37^{\circ}\pm0.008^{\circ}$ and $\rm{DEC}_{J2000}=-6.92^{\circ}\pm0.008^{\circ}$ with a size of $0.22^{\circ}\pm0.01^{\circ}$. These observations also revealed a second source located to the South of \hess, when considering the International Celestial Reference System (ICRS). No official name was attributed to this potential new source. However, no significant emission from the \snr\ region was claimed. Recent results from the new HGPS \citep{2018A&A...612A...1H} characterise the region of \hess\ as a superposition of three Gaussian sources with a total extension of $0.36^{\circ}\pm 0.03^{\circ}$. 
This region of the sky was also covered by HAWC at energies above 1\,TeV. The second HAWC catalog \citep{2017ApJ...843...40A} shows a 15$\sigma$-excess compatible with the position of \hess\ after $1.5$ year observation time.

In this paper, we study the interesting region centered around SNR
G24.7+0.6 with \fermi-LAT in the 
energy range between 60\,MeV and 500\,GeV. We also explore with the MAGIC telescopes the region around it to investigate 
the spectral behaviour above 150\,GeV in order to constrain the emission region observed by \fermi-LAT 
around the SNR.

\section{Observations and data analysis}

\subsection{\textit{Fermi}-LAT}  \label{Fermi}

We analysed $\sim$8 years of data spanning from 4 August 2008 to 13 June 2016 with energies between 60\,MeV and 500\,GeV. The dataset was analysed using \textit{Fermipy}\footnote{http://fermipy.readthedocs.io/en/latest/} v0.13.3: a set of python programmed tools that automatise the PASS8 analysis with the Fermi Science Tools\footnote{https://fermi.gsfc.nasa.gov/ssc/data/analysis/documentation/}. The CLEAN event class was chosen for this analysis since the source is shown to be extended \citep{2016ApJS..224....8A}. In addition, it benefits from a lower background above 3\,GeV with respect to the SOURCE event class. We used P8R2\_CLEAN\_V6 instrument response function (IRF). This IRF is divided into three event types, FRONT/BACK, PSF and EDISP. For the analysis presented here we used the PSF partition which guarantees the best quality of the reconstructed direction. This PSF partition is subdivided into four quartiles increasing in quality from PSF0 to PSF3, each of which with its zenith angle cut to reduce the background from the Earth limb. Thus photons with zenith angles larger than 70, 75, 85 and 90 for PSFs ranging from the worst to the best were excluded. The analysis of the four quartiles was performed independently and combined in later stages of the analysis by means of a joint likelihood fit.\footnote{http://fermipy.readthedocs.io/en/latest/fitting.html}

We performed a maximum likelihood analysis in a circular region of 20\degr\ 
radius centered on the radio source position $\rm{RA}_{J2000}=278.57$\degr; $\rm{DEC}_{J2000}=-7.09$\degr , this region will be referred as the region of interest (ROI). The emission model for our ROI includes the LAT sources listed in the third LAT catalog (3FGL, \citealt{2015ApJS..218...23A}) within a region of 30\degr\ 
radius around \snr\ and the diffuse $\gamma$-ray background models; the Galactic diffuse emission modelled by \textit{gll\_iem\_v06.fits} 
and the isotropic component by \textit{iso\_P8R2\_CLEAN\_V6\_PSFX\_v06.txt} (where X identifies the number of the PSF 
quartile), including the instrumental background and the extragalactic radiation. 
Sources lying within 4\degr\ from the source of interest were fit with all their spectral parameters left free. For sources between 4\degr\ and 7\degr\ and the Galactic diffuse and isotropic components, only the normalisation parameters were allowed to vary. All the spectral parameters for sources located farther than 7\degr\ from the source of study remained fixed in the maximum likelihood fit. 

Due to strong contamination from diffuse emission in the
Galactic plane at low energies and the large PSF, both mainly below 1\,GeV, in order to study the morphology of the 
source we performed a specific analysis to the LAT data above 1\,GeV in a 8$^{\circ} \times $8$^{\circ}$ region centered on 
the \snr\ radio position. 
Given that our source of interest might be associated with two 3FGL sources (3FGL\,J1834.6--0659 and 3FGL\,J1833.9--0711), 
which are tagged as `confused', meaning that they can arise from a wrongly 
modeled background or a confused source pile-up, we removed them from the model to study in more detail the residual map. We found that 
replacing these sources with a single point-like source (we called it FGES\,J1834.1--0706 as in \citealt{2017ApJ...843..139A}) 
located at the radio position increases the likelihood value.
We performed a \emph{localisation procedure}\footnote{http://fermipy.readthedocs.io/en/latest/advanced/localization.html} 
within a region of 3$^{\circ} \times $3$^{\circ}$ to determine the correct position of FGES\,J1834.1--0706 and we tested for a possible 
extended morphology modeling our source with a Gaussian function rather than a point-like source.
Assuming a power-law 
spectral shape with spectral index -2, we performed an iterative likelihood fit for values of the source 
extension\footnote{http://fermipy.readthedocs.io/en/latest/advanced/extension.html}  ranging from 0.01$^{\circ}$\ to 
1.01$^{\circ}$\ with a step of 0.1$^{\circ}$.

To obtain the spectral energy distribution, we split the
60\,MeV--500\,GeV energy range into 10 logarithmically spaced
bins. Each spectral point has at least a TS value greater than or
equal to 4, otherwise 95\% confidence level (CL) flux ULs were
computed.

\subsection{MAGIC telescopes}  \label{MAGIC}

The VHE $\gamma$-ray observations of \snr\ were performed using the MAGIC telescopes. 
MAGIC observed \snr\ between April 5th and August 29th, 2014, for a total of 33 hours, at zenith angles between $35\degr$ and $50\degr$, yielding an analysis energy threshold of $\sim$200\,GeV. The observations were performed in wobble-mode \citep{1994APh.....2..137F} at four symmetrical positions 0.4\degr\ away from the source, so that the background can be estimated simultaneously. After quality cuts, which account for hardware problems, unusual background rates and reduced atmospheric transparency, 31 hours of high quality data were selected. 

The analysis of the MAGIC data was performed using the standard MAGIC Analysis and Reconstruction Software, MARS \citep{2010ascl.soft11004M, Zanin2013ICRC}. 
In particular, we derived On-maps of $\gamma$-like events based on their arrival directions in sky coordinates. ON-maps need a reliable background determination in order to minimize the contribution of hadronic cosmic rays surviving data selection cuts. To reconstruct the background maps from wobble observations we use the Exclusion Map technique implemented in SkyPrism \citep{skyprism}. The Exclusion Map technique allows to estimate the background with no need of prior knowledge of the position of the source under evaluation while we exclude from the computation regions 
containing known sources.
ON and Background maps are used as input files for a two-dimensional maximum likelihood fit of the source model that is performed 
using the \emph{Sherpa} package \citep{2007ASPC..376..543D,2001SPIE.4477...76F}. 
Specifically, the source model is constructed and optimized by using
an iterative method in a likelihood approach. First, a single
Gaussian-shaped source is added to a model containing only the
isotropic background. Different positions and extensions of the source
are evaluated and the values maximising the likelihood value are
assigned to the source. A second Gaussian-shaped source is added
to the model and the same procedure is executed; positions and
extensions for both sources are re-calculated. These two nested models
are compared through their maximum likelihood fit value. Additional
Gaussian-shaped sources are iteratively introduced to the model
until the maximum likelihood fit is no longer improved. Given the
  complexity of the region, together with the drop in the sensitivity
  of MAGIC  when one moves away from the center, only symmetric
  gaussian-type sources could be tested.  For the spectral analysis
of the best-fit  model obtained, we performed an additional
one-dimensional maximum likelihood fit using \texttt{SkyPrism}.

\begin{figure*}
  \centering
  \includegraphics[width=0.49\textwidth]{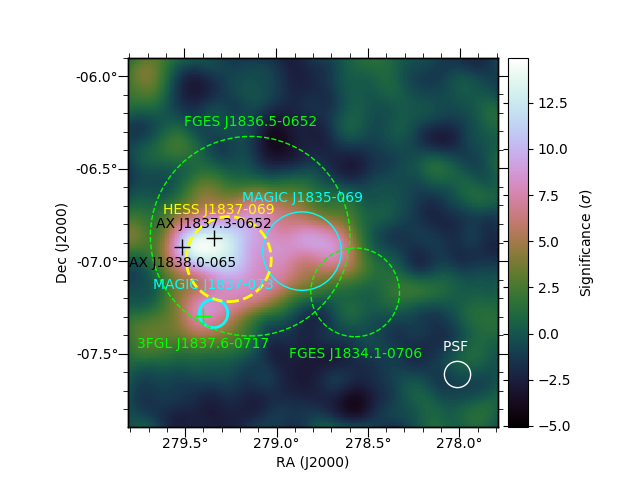} 
  \includegraphics[width=0.49\textwidth]{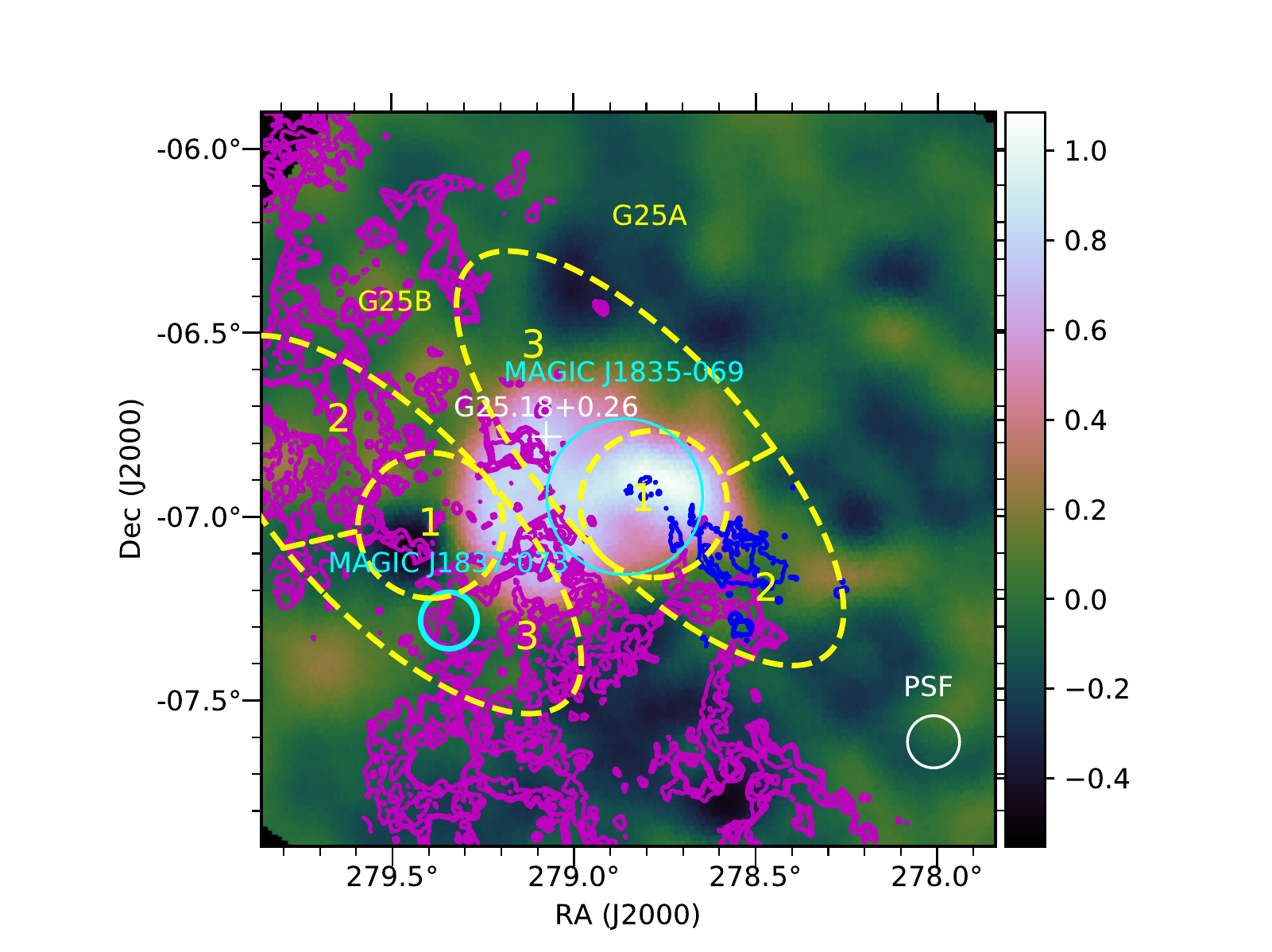}
  \caption{
    \textit{Left:} $2^{\circ} \times 2^{\circ}$ significance map of
    the region obtained with MAGIC. The extension of \mgc\ and \mgcS\
    are represented by the thin and thick blue circles, respectively,
    while \fermi-LAT sources from FGES and 3FGL catalogs in the FoV
    are displayed by green dashed lines and a cross. The position and
    extension of \hess\ as measured in this work are displayed by a
    yellow dashed circle. The positions of the two X-ray PWN
      candidates in \citet{2008ApJ...681..515G} are marked with black
      crosses.  AX\,J1838.0--0655 is proposed as counterpart of \hess.
  \textit{Right:} Residual map (data-model) in counts normalised to 1 derived from MAGIC data after subtracting the emission from \hess\ and \mgcS.
Over the MAGIC map, the \snr\ radio emission and $^{13}$CO contours are
overlaid  in blue and magenta, respectively. The integrated in all
velocities $^{13}$CO (J=1-0) contours from the Galactic Ring Survey are selected from 7\,K to 13\,K in step of size 3 to emphasise the cloud spacial distribution. The yellow dashed ellipses (G25A and G25B) along with their three components represent the \fermi-LAT sources found within the region by \citet{2017ApJ...839..129K}. 
The white cross displays the position of the OB association/cluster G25.18+0.26 identified through X-ray observation by \citet{2017ApJ...839..129K}.}
  \label{fig:map}
\end{figure*}


\section{Results}	

The obtained MAGIC significance skymap, shown in Figure~\ref{fig:map} (\textit{left} panel) in the ICRS coordinate system, shows significant extended emission at energies larger than 200\,GeV. The two-dimensional likelihood morphological analysis led to the detection of three distinct sources:

\begin{itemize}
\item
The brightest source is identified with \hess\ \citep{2006ApJ...636..777A, 2008AIPC.1085..320M}. It presents an extended morphology characterized by a symmetric Gaussian of $0.23^{\circ}\pm0.01^{\circ}$ size centered at RA$_{J2000}$=$279.26^{\circ}\pm0.02^{\circ}$ and DEC$_{J2000}$=$-6.99^{\circ}\pm0.01^{\circ}$.
This emission is also associated with the extended source FGES\,J1836.5--0652 in the \textit{Fermi}-LAT catalog of extended sources in the Galactic plane. 

\item
The excess to the South from \hess\ is significantly detected
  with a peak significance of 8.2$\sigma$. We added a new point-like
  source in our morphological model of the region to account for this
  excess. The two-sources model is favoured with respect to the
  one-single-source one at 7.7$\sigma$ level. 
This new source is well-fit with a Gaussian shape with an 
extension of $0.08^{\circ}\pm0.05^{\circ}$ centered at $\rm{RA}_{J2000}=279.34^{\circ}\pm0.14^{\circ}$ and $\rm{DEC}
_{J2000}=-7.28^{\circ}\pm0.24^{\circ}$. The comparison between a model
describing the source as point like and a model treating the source as
a Gaussian results in a $\Delta$TS of 23 or $\sim$5$\sigma$, favouring the second model to describe the total emission. Spatially coincident with the hotspot reported in \citet{2008AIPC.1085..320M}, we named it \mgcS\ since no name was officially previously attributed to it. This source is also coincident with 
3FGL\,J1837.6--0717 reported in the Third Catalog of \textit{Fermi}-LAT sources \citep{2015ApJS..218...23A}. 

\item
The third significant source (with a peak significance of 11.0$\sigma$) is, for the first time, detected at VHE, and it is named \mgc. The source is resolved at a level of 13.5$\sigma$ when adding it to the global fit. It is significantly extended and well modelled by a Gaussian of $0.21^{\circ}\pm 0.05^{\circ}$ centered at $\rm{RA}_{J2000}
=278.86^{\circ}\pm0.23^{\circ}$ and $\rm{DEC}_{J2000}=-6.94^{\circ}\pm0.05^{\circ}$. The extended nature of the source is favoured at a level of 7.1$\sigma$. Its center position is offset by 
$0.34^{\circ}$ with respect to the center of the \snr. In particular, it lies between two extended sources 
detected above 10 GeV by \textit{Fermi}-LAT, FGES\,J1836.5--0652 and the FGES\,J1834.1-- 0706, being the first associated 
to \hess\ and the second to the \snr.  
\end{itemize}

\begin{figure*}
  \centering
  \includegraphics[width=0.49\textwidth]{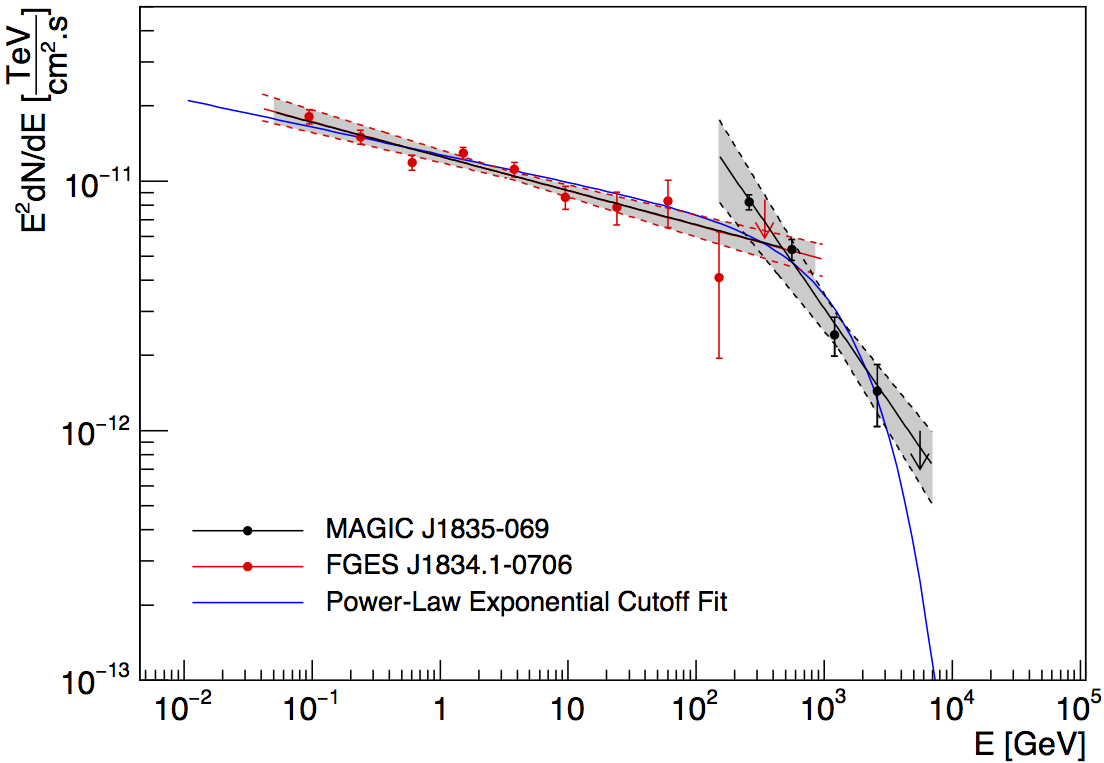}
  \includegraphics[width=0.49\textwidth]{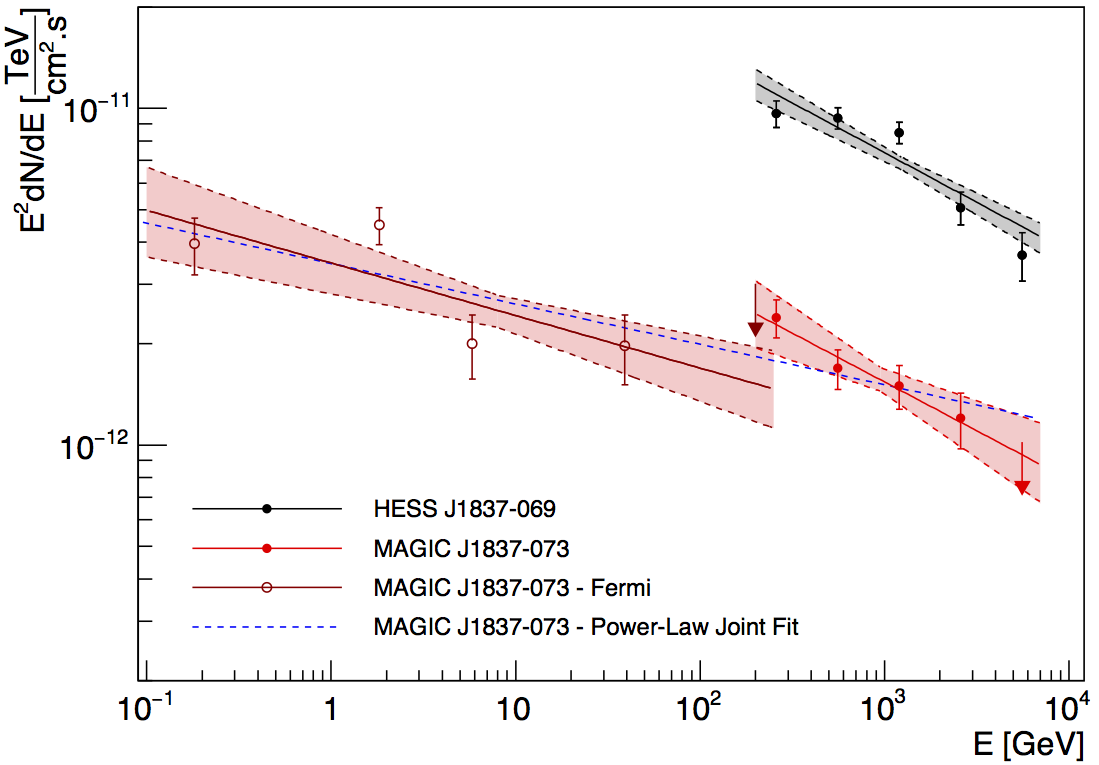}
  \caption{ {\it{Left:}} Spectral energy distribution of FGES\,J1834.1--0706 (red circles) and \mgc\ (black circles) between 60\,MeV and 10\,TeV obtained with the analysis described in Section 2.2. In the \fermi\ energy range the spectrum follows a power-law of index 2.14 while it softens in the MAGIC range to an index of 2.74. The EPWL fit for the whole energy range is represented by a blue line. Light gray bands are the statistical uncertainties. {\it{Right:}} Spectral energy distribution of \hess\ (black) and \mgcS\ (red), measured by MAGIC between 200\,GeV and 10\,TeV. Solid lines represent the power-law fits applied to each spectrum. Light shaded bands are the statistical uncertainties. The spectrum measured for \mgcS\ with \fermi-LAT along with its power-law fit is represented in dark red. Blue dashed line represents the joint $\chi^2$ fit of \mgcS\ between 60 MeV and 10 TeV.} 
  \label{fig:spectrumG24}
\end{figure*}

Figure \ref{fig:spectrumG24} shows the SED obtained for the three sources with the likelihood method explained in Section 2.2 and using the above-described morphologies as extraction regions. The spectral fit parameters are summarized in Table \ref{tab:sed}. 
The differential energy spectrum of \hess\ is well represented by a power-law function with a photon index of $2.29\pm0.04$ and an integral flux above 200 GeV of $(7.2\pm0.3)\times10^{-11}$\,erg\,cm$^{-2}$\,s$^{-1}$. The spectrum obtained is compatible within statistical errors with those measured by H.E.S.S., $2.27\pm0.06$ in \citet{2006ApJ...636..777A} and $2.34\pm0.04$ in \citet{2008AIPC.1085..320M}.
For \mgcS, the best spectral fit model is a power-law with a $2.29 \pm 0.09$ photon index and an integral flux above 200 GeV of $(1.5\pm0.1)\times10^{-11}$\,erg\,cm$^{-2}$\,s$^{-1}$. The emission fades away above 3\,TeV, and the calculated 95\% CL UL at 6\,TeV does not constrain any potential cut-off.
Finally, the energy spectrum of \mgc\ is best fit by a power-law function with a photon index of $2.74 \pm 0.08$ and an integral flux above 200 GeV of $(4.4\pm0.6)\times10^{-11}$\,erg\,cm$^{-2}$\,s$^{-1}$.

The results obtained with our \textit{Fermi}-LAT analysis are in good agreement with previously published ones. Two sources 
are detected in the surrounding of \snr; FGES\,J1834.1--0706 and the counterpart of the 
MAGIC source \mgcS, 3FGL\,J1837.6--0717. The first shows an extended Gaussian emission of $0.24^{\circ}\pm 
0.01^{\circ}$ centered at $\rm{RA}_{J2000}=278.57^{\circ}\pm0.01^{\circ}$ and $\rm{DEC}_{J2000}=-7.19^{\circ}
\pm0.02^{\circ}$, offset by 0.1$^{\circ}$ from the radio position. The significance of the extension is of 11.4$\sigma$ ($\rm{TS}
_{ext} = 131$\footnote{It was calculated from $\rm{TS}_{ext} = \rm{TS}_{gauss} - \rm{TS}_{point}$ as stated in \cite{2012ApJ...756....5L}}). This result is in agreement with the one published in the FGES catalog. As stated in Section 1, we consider as 
reference analysis the one of the FGES catalog, thus we refer to the source found in our analysis as FGES\,J1834.1-- 0706.

The energy spectra obtained with our \textit{Fermi}-LAT analysis from 60 MeV to 
500 GeV for FGES\,J1834.1-- 0706 and \mgcS\ are represented in Figure~\ref{fig:spectrumG24}. \mgcS, for which we used the morphology derived in the MAGIC analysis, 
exhibits a power-law spectrum with a photon index of $\Gamma=(2.15 \pm 0.05)$ and a normalisation factor of $N_0=(3.9\pm 
0.4) \times 10^{-8}$\,TeV$^{-1}$\,cm$^{-2}$\,s$^{-1}$ at the decorrelation energy of 8\,GeV. The mismatch between the flux 
level obtained by the two instruments is well within the systematic uncertainties, estimated to be of the order of 15\% for MAGIC. 
A joint $\chi^2$ fit of \mgcS\ between 60 MeV and 10 TeV results in a similar power-law of photon index $\Gamma_{joint}
=(2.12 \pm 0.02)$ with a factor of $N_0=(1.52\pm 0.1) \times 10^{-12}$\,TeV$^{-1}$\,cm$^{-2}$\,s$^{-1}$ at 1\,TeV.
On the other hand, FGES\,J1834.1-- 0706 shows a power-law spectrum with a photon index of $\Gamma=(2.14 \pm 0.02)$ and a normalisation 
factor of $N_0=(2.9\pm 0.1) \times 10^{-7}$\,TeV$^{-1}$\,cm$^{-2}$\,s$^{-1}$ at the decorrelation energy of 5.8\,GeV. In this 
case, the energy spectrum of FGES\,J1834.1-- 0706 connects smoothly with that of \mgc\ even though the extraction regions 
are not exactly the same, thus suggesting that the two sources most likely have a common origin. Under this assumption, we 
performed a joint $\chi^2$ fit between 60 MeV and 10 TeV that resulted in a power-law function with an exponential cut-off 
(hereafter, EPWL), $F_0 \left(\frac{E}{E_0}\right)^{-\Gamma}{e^{-\frac{E}{E_c}}}$, where $F_0$ is the prefactor; $E_0$ is the 
decorrelation energy; $E_c$ is the cut-off energy and $\Gamma$ is the photon index. The resulting fitting parameters are 
provided in Table \ref{tab:fit}.

\begin{table}
\centering
\caption{Fitting spectral parameters of the three sources detected by MAGIC. $\Gamma$ is the photon index, and F$_0$ the normalisation factor at the decorrelation energy E$_0$. }
\label{tab:sed}
\begin{tabular}{cccc}
\hline
& F$_0$ & $\Gamma$ & E$_0$  \\
& [TeV$^{-1}$ cm$^{-2}$ s$^{-1}$] & & [TeV] \\
\hline
\hess & $(4.4\pm0.2)\times10^{-12}$ & $2.29\pm0.04$ & 1.25 \\
\mgcS & $(1.7\pm0.1)\times10^{-12}$ & $2.29\pm0.09$ & 0.95 \\
\mgc & $(1.4\pm0.2)\times10^{-12}$ & $2.74\pm0.08$ & 1.31 \\
\hline
\end{tabular}
\end{table}

\begin{table} 
\caption{Joint $\chi^2$ fit spectral parameters for \snr\ from 60\,MeV to $\sim$10\,TeV.  Photon index, $\Gamma$, normalisation factor F$_0$ at the decorrelation energy E$_0$ and cutoff energy are presented. }
\label{tab:fit}
\begin{tabular}{ccccccc}
\hline
& F$_0$ & $\Gamma$ & E$_C$ & E$_0$ \\
& [TeV$^{-1}$ cm$^{-2}$ s$^{-1}$] & & [TeV] & [GeV]\\
\hline
EPWL & $(9.1\pm3.0)\times10^{-10}$ & $2.12\pm0.02$ & $1.9\pm0.5$ & 92 \\ 
\hline
\end{tabular}
\end{table}


\section{Discussion}

We observed the FoV of \snr\ with the MAGIC telescopes, following the detection of a hard-spectrum source reported by the 
LAT collaboration \citep{2016ApJS..222....5A}, coincident with the position of the remnant. The analysis of 31 hours of data 
using the \emph{Sherpa} package on the reconstructed skymap resulted in the detection of three different sources in the 
MAGIC data set. The brightest one, located at RA$_{J2000}$=$279.26^{\circ}\pm0.02^{\circ}$ and DEC$_{J2000}$=$-6.99^{\circ}\pm0.01^{\circ}$, has 
been previously reported  by the H.E.S.S. collaboration and named\hess. It is classified as PWNe based on the detailed spectral-morphological study performed by the H.E.S.S. collaboration and the discovery of the associated X-ray source AX\,J1838.0--0655 \citep{2008ApJ...681..515G}. The spectral features derived by MAGIC in this 
region are compatible within errors with those reported by H.E.S.S. 

To the South, \mgcS, a $\gamma$-ray 
excess located $0.34^{\circ}$ away from \hess\ is detected at a level of 7.7$\sigma$. The spectrum of this source extends to low energies. The origin of this emission remains unclear, although, under a first approximation assumption of one single parent population, an hadronic scenario is most likely to explain a single power-law spectrum up to few tens of TeV. The region was subject of observations with \textit{XMM}-Newton \citep{2017ApJ...839..129K} in a search for a multi-wavelength counterpart of the GeV emission they detect (G25B in Figure \ref{fig:map}, \textit{right} panel). 
No PWN, SNR, or pulsar with spin-down luminosity $>1 \times
10^{34}$\,erg\,s$^{-1}$ was found in the region. However, the region
is rich in molecular content at velocities $v = 45 -
65$\,km\,s$^{-1}$. In the GeV regime, it has been postulated as
  possible association with a bubble identified with the stellar
  cluster candidate G25.18+0.26, but no sign of such connection can be
  derived from the TeV data. If, based on the spectral shape and the
  presence of molecular target, we assume an hadronic origin of the
  emission \citep{2017ApJ...839..129K}. The total luminosity of \mgcS\
  above 100 MeV will amount to $L_{\gamma} = 1.3
  \times10^{35}$\,erg\,s$^{-1}$, for a distance of $d = 5$\,kpc. This implies a total cosmic rays energy of $W_{p} \approx 2.1 \times 10^{50}$\,erg$\left(\frac{\rm{cm}^{-3}}{n}\right)$, being $n$ the ambient proton density. This number is comparable to the ones found in other clusters such Westerlund 2 \citep{2018A&A...611A..77Y} or Cygnus Cocoon \citep{2011Sci...334.1103A}. Such large luminosity could be achieved by assuming a quasi-continuous injection of cosmic rays, powered by the kinetic energy released for instance in the winds of massive stars ($\sim1\times10^{38}$\,erg\,s$^{-1}$), integrating during the cluster lifetime (typically $\sim1\times10^{4}$\,yrs).

\begin{figure}
  \centering
  \includegraphics[width=0.5\textwidth]{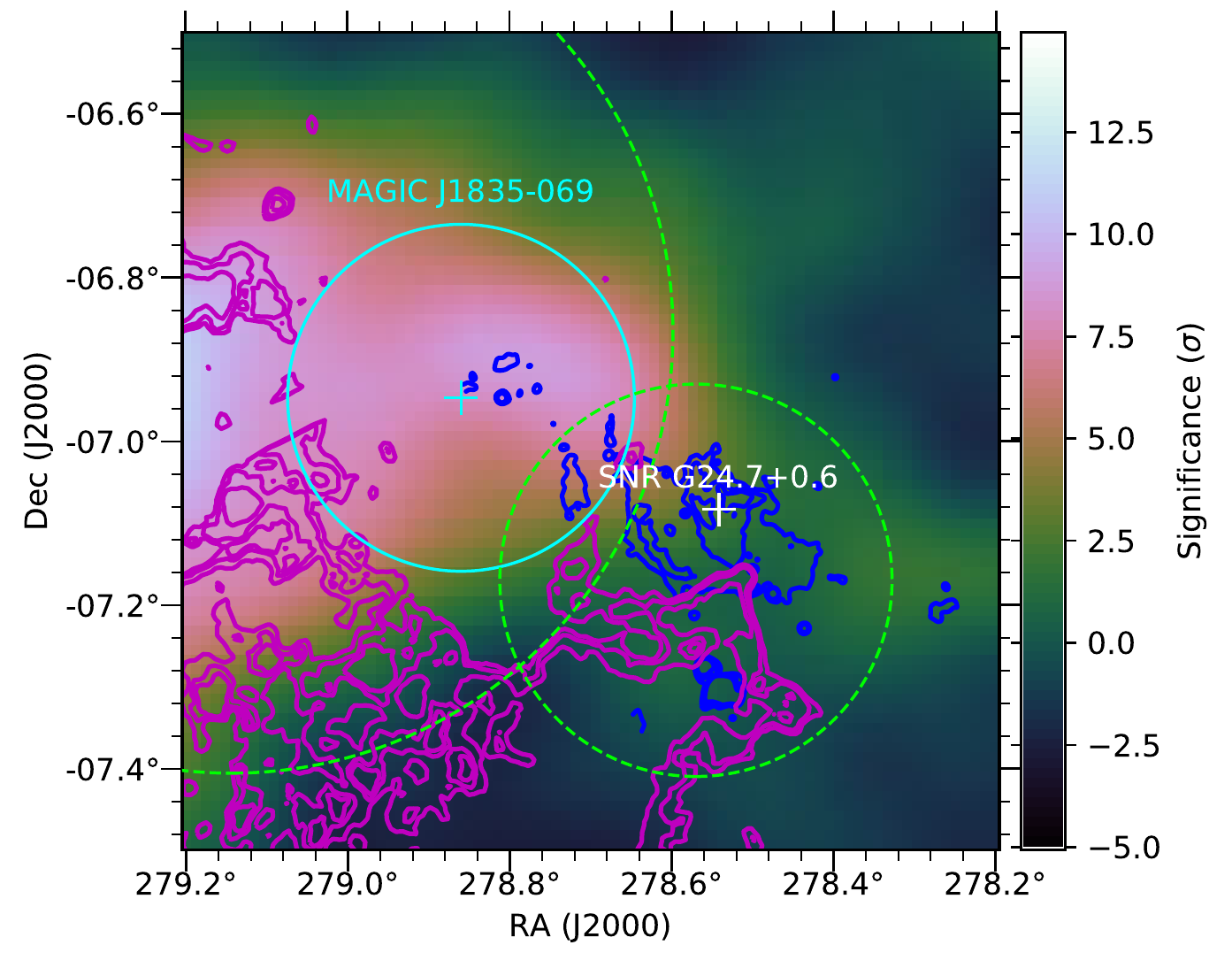} 
  \caption{
    $1^{\circ} \times 1^{\circ}$ significance map of the region
    obtained with MAGIC. \mgc\ is marked with a blue line. The
      green circles show the extension of the \fermi-LAT sources from
      FGES catalog. The VLA radio contours of the region are overlaid
      in blue, showing the extension of \snr\, centered at the
      position of the white cross. The integrated $^{13}$CO $(J=1\rightarrow0)$ intensity
      contours from the Galactic Ring Survey are showed in magenta\protect\footnotemark[7].}
  \label{fig:map2}
\end{figure}

Finally, the statistical test performed allows to resolve \mgc\ ($\rm{RA}_{J2000}=278.86^{\circ}\pm0.23^{\circ}$; $\rm{DEC}_{J2000}=-6.94^{\circ}\pm0.05^{\circ}
$) from \hess\ at a 13.5$\sigma$ level. Moreover, the projected distance of the new $\gamma$-ray enhancement to the pulsar 
associated to \hess\ (for a distance of 6.6 kpc, from \citet{2008ApJ...681..515G}), is more than $\sim$65 pc, which, if not impossible, 
makes the association between the two sources unlikely. \mgc, however, partially overlaps with the emission detected with 
\textit{Fermi}-LAT (see Figure \ref{fig:map} and Figure \ref{fig:map2} for a zoom in of the region). Indeed, a new analysis presented by \cite{2017ApJ...843..139A} describes the complex region with three very extended 
sources, being the MAGIC source comprised between two sources; FGES\,J1836.5--0652, which includes also \hess, 
and FGES\,J1834.1--0706 which is consistent with 3FHL\,J1834.1--0706e on the position of the \snr. The flux measured with 
MAGIC is in good agreement with the one measured by LAT, extending the spectrum from 60 MeV to 10 TeV with a spectral 
photon index of $\sim$2.74. The VHE broad band spectral shape shows a clear break in the GeV-TeV regime. This change of 
slope can be described by a power-law with an exponential cut-off at E$_{\rm C}=1.9$\,TeV.  The source shows an extended 
morphology and it is offset 0.34$^{\circ}$ with respect to center of the remnant, in a region where the later seems to be blowing 
an IR shell. The measured offset translates onto a projected size of 30\,pc at the distance of 5\,kpc. The CO-rich 
surrounding of \snr\ could be originating the detected GeV-TeV emission, and the offset between the emission detected by LAT 
and the MAGIC source could be interpreted in terms of diffusion mechanism similar to what was proposed for IC 443 
\citep{2008MNRAS.387L..59T,2010MNRAS.408.1257T}, since the diffusion
radius of runaway protons of 100\,GeV could account for this
distance. However the large error in the position and the complexity
of the region in the GeV and TeV regime prevent further conclusions in
that sense. Nevertheless, in this scenario and similarly to other
evolved SNR, the VHE LAT/MAGIC combined spectrum model can be
explained as a result of proton-proton interaction between the cosmic
rays accelerated in \snr\ and those in the surrounding gas. The total
luminosity above 100\,GeV amounts
$L_{\gamma}=7.5\times10^{34}$\,erg\,s$^{-1}$ , which translates to a
total energetics stored in accelerated protons of
$W_{p}=1.3\times10^{50}$\,erg$\left(\frac{\rm{cm}^{-3}}{n}\right)$.

\footnotetext[7]{obtained from \url{https://www.bu.edu/galacticring/new_data.html}}

A second scenario involving a yet-undiscovered PWN associated to the remnant cannot be discarded. At 
a distance of \textit{d}$\sim$5~kpc, the separation between \mgc\ and the position of the remnant is within the range of offsets found 
in VHE PWNe (see Figure 6 from \citealt{2018A&A...612A...2H}). The corresponding surface brightness, in the energy range 
from 1 to 10\,TeV, would be $\sim1.2\times10^{30}$\,erg\,s$^{-1}$\,pc$^{-2}$. Applying the correlation found by 
\cite{2018A&A...612A...2H} (S$\sim\dot{\rm E}^{0.81\pm0.14}$), an extremely bright $\dot{\rm E}
\sim1.4\times10^{37}$erg\,s$^{-1}$ pulsar should be powering the VHE source. Both the upper and the lower limit of the spin-down luminosity (S$\sim\dot{\rm E}^{0.67}$ and S$\sim\dot{\rm E}^{0.95}$, respectively) seem unrealistically large for not being detected either in $\gamma$-ray or radio. However, the strong confusion 
due to the several extended sources in the field limits the detection of such pulsars in the GeV regime. In addition, the 
extension of the PWN would exceed the SNR size, rendering this scenario unlikely if the putative PWN is connected to the 
SNR.

Recently, \citet{2017ApJ...839..129K} carried out a study of the $\gamma$-ray emission coming from the region around, $
\rm{RA}_{J2000}=279.22^{\circ}$ and $\rm{DEC}_{J2000}=-7.05^{\circ}$, with the \fermi-LAT telescope. They found that the 
emission detected is divided into two elliptical extended region, G25A and G25B, composed of 3 components each (see Figure 
\ref{fig:map}, \textit{right} panel). For G25A, all three components have the same spectral shapes while for G25B, the G25B1 component has a harder 
spectrum than the other two. Due to their elongated morphology and spectral similarity (similar surface brightness and hard 
energy spectra; $\Gamma=(2.14 \pm 0.02)$ and $\Gamma=(2.11 \pm 0.04)$, respectively), they suggested that both $
\gamma$-ray emissions are produced by the same astrophysical object. In addition, through X-ray observations of the region 
with \textit{XMM}-Newton they found the candidate young massive OB association/cluster, G25.18+0.26 (Figure 
\ref{fig:map}, \textit{right} panel). They proposed that both extended $\gamma$-ray emissions (G25A and G25B) are associated with an star 
forming region driven by G25.18+0.26. 
Assuming the scenario proposed by \citet{2017ApJ...839..129K} in which either the accelerated particles are interacting with 
regions of enhanced gas density or particles are being accelerated within these regions, current TeV telescopes like MAGIC 
should reveal a diffuse $\gamma$-ray emission from the whole G25A and G25B regions. However, as seen from the maps, 
MAGIC only detects emission from the G25A1 component that is coincident with \mgc. We can conclude it is unlikely that the 
emission detected at VHE with MAGIC comes from the OB association/cluster G25.18+0.26 detected in X-rays.


\section{Conclusions}

MAGIC observations of the field of view of the \snr\ resulted in the
discovery of a new TeV source in the Galactic plane, \mgc, detected
above $\sim$150\,GeV. The position of \mgc\ is compatible  at
1.5$\sigma$ with the center of SNR, which is in turn associated with
the \fermi\ source FGES J1834.1--0706. Based on the good agreement
between the LAT and MAGIC spectral measurements, the two sources are
likely to be associated. The link with the SNR is also plausible if
one consider the diffusion radius of particles to explain the observed
offset. The GeV-TeV emission observed by \fermi\ and MAGIC can be
interpreted as cosmic rays accelerated within the remnant interacting
via proton-proton collisions with the $^{13}$CO surrounding medium.

A second statistically significant detection of a slightly extended
$\gamma$-ray signal from the south of \hess\ is reported. The spectrum
of the source extends to 3\,TeV with no sign of an spectral
break. Although the PWN scenario cannot
be ruled out, this detection is believed to be produced by cosmic rays
interacting with a stellar cluster. If the latter is confirmed, \mgcS\
will be part of the scarcely populated group of similar objects like
Westerlund 1 and 2 or the Cygnus cocoon and may contribute to a better
understanding of whether these objects can account for the Galactic
cosmic ray flux.

\section*{Acknowledgements}
%
%
We would like to thank the Instituto de Astrof\'{\i}sica de Canarias for the excellent working conditions at the Observatorio del Roque de los Muchachos in La Palma. The financial support of the German BMBF and MPG, the Italian INFN and INAF, the Swiss National Fund SNF, the ERDF under the Spanish MINECO (FPA2015-69818-P, FPA2012-36668, FPA2015-68378-P, FPA2015-69210-C6-2-R, FPA2015-69210-C6-4-R, FPA2015-69210-C6-6-R, AYA2015-71042-P, AYA2016-76012-C3-1-P, ESP2015-71662-C2-2-P, CSD2009-00064), and the Japanese JSPS and MEXT is gratefully acknowledged. This work was also supported by the Spanish Centro de Excelencia ``Severo Ochoa'' SEV-2012-0234 and SEV-2015-0548, and Unidad de Excelencia ``Mar\'{\i}a de Maeztu'' MDM-2014-0369, by the Croatian Science Foundation (HrZZ) Project IP-2016-06-9782 and the University of Rijeka Project 13.12.1.3.02, by the DFG Collaborative Research Centers SFB823/C4 and SFB876/C3, the Polish National Research Centre grant UMO-2016/22/M/ST9/00382 and by the Brazilian MCTIC, CNPq and FAPERJ. 
 
\label{lastpage}

\bibliography{biblio}
\vspace{5mm}
\noindent$^{1}$ {Inst. de Astrof\'isica de Canarias, E-38200 La Laguna, and Universidad de La Laguna, Dpto. Astrof\'isica, E-38206 La Laguna, Tenerife, Spain} \\
$^{2}$ {Universit\`a di Udine, and INFN Trieste, I-33100 Udine, Italy} \\
$^{3}$ {National Institute for Astrophysics (INAF), I-00136 Rome, Italy} \\
$^{4}$ {ETH Zurich, CH-8093 Zurich, Switzerland} \\
$^{5}$ {Universit\`a di Padova and INFN, I-35131 Padova, Italy} \\
$^{6}$ {Technische Universit\"at Dortmund, D-44221 Dortmund, Germany} \\
$^{7}$ {Croatian MAGIC Consortium: University of Rijeka, 51000 Rijeka, University of Split - FESB, 21000 Split, University of Zagreb - FER, 10000 Zagreb, University of Osijek, 31000 Osijek and Rudjer Boskovic Institute, 10000 Zagreb, Croatia.} \\
$^{8}$ {Saha Institute of Nuclear Physics, HBNI, 1/AF Bidhannagar, Salt Lake, Sector-1, Kolkata 700064, India} \\
$^{9}$ {Max-Planck-Institut f\"ur Physik, D-80805 M\"unchen, Germany} \\
$^{10}$ {now at Centro Brasileiro de Pesquisas F\'isicas (CBPF), 22290-180 URCA, Rio de Janeiro (RJ), Brasil} \\
$^{11}$ {Unidad de Part\'iculas y Cosmolog\'ia (UPARCOS), Universidad Complutense, E-28040 Madrid, Spain} \\
$^{12}$ {University of \L\'od\'z, Department of Astrophysics, PL-90236 \L\'od\'z, Poland} \\
$^{13}$ {Deutsches Elektronen-Synchrotron (DESY), D-15738 Zeuthen, Germany} \\
$^{14}$ {Institut de F\'isica d'Altes Energies (IFAE), The Barcelona Institute of Science and Technology (BIST), E-08193 Bellaterra (Barcelona), Spain} \\
$^{15}$ {Universit\`a di Siena and INFN Pisa, I-53100 Siena, Italy} \\
$^{16}$ {Universit\`a di Pisa, and INFN Pisa, I-56126 Pisa, Italy} \\
$^{17}$ {Universit\"at W\"urzburg, D-97074 W\"urzburg, Germany} \\
$^{18}$ {Finnish MAGIC Consortium: Tuorla Observatory and Finnish Centre of Astronomy with ESO (FINCA), University of Turku, Vaisalantie 20, FI-21500 Piikki\"o, Astronomy Division, University of Oulu, FIN-90014 University of Oulu, Finland} \\
$^{19}$ {Departament de F\'isica, and CERES-IEEC, Universitat Aut\'onoma de Barcelona, E-08193 Bellaterra, Spain} \\
$^{20}$ {Universitat de Barcelona, ICC, IEEC-UB, E-08028 Barcelona, Spain, E-mail: \texttt{dgalindo@fqa.ub.edu}} \\
$^{21}$ {Japanese MAGIC Consortium: ICRR, The University of Tokyo, 277-8582 Chiba, Japan; Department of Physics, Kyoto University, 606-8502 Kyoto, Japan; Tokai University, 259-1292 Kanagawa, Japan; RIKEN, 351-0198 Saitama, Japan} \\
$^{22}$ {Inst. for Nucl. Research and Nucl. Energy, Bulgarian Academy of Sciences, BG-1784 Sofia, Bulgaria} \\
$^{23}$ {Humboldt University of Berlin, Institut f\"ur Physik D-12489 Berlin Germany} \\
$^{24}$ {also at Dipartimento di Fisica, Universit\`a di Trieste, I-34127 Trieste, Italy}\\
$^{25}$ {also at Port d'Informaci\'o Cient\'ifica (PIC) E-08193 Bellaterra (Barcelona) Spain} \\
$^{26}$ {also at INAF-Trieste and Dept. of Physics \& Astronomy, University of Bologna}\\
$^{27}$ {Institute for Space Sciences (ICE, CSIC), E-08193 Barcelona, Spain, E-mail: \texttt{wilhelmi@ice.csic.es}} \\
$^{28}$ {Institut d'Estudis Espacials de Catalunya (IEEC), 08034 Barcelona, Spain}\\
$^{29}$ {AvH Guest at Deutsches Elektronen Synchrotron DESY, 15738 Zeuthen, Germany}\\
$^{30}${Instituci\'o Catalana de Recerca i Estudis Avancats (ICREA), E-08010 Barcelona, Spain}\\
$^{31}$ {Max-Planck-Institut f\"ur Kernphysik, D-69029 Heidelberg, Germany, E-mail: \texttt{roberta.zanin@mpi-hd.mpg.de}} \\

\end{document}